\documentclass[aps,showpacs,twocolumn]{revtex4}
\usepackage{amsmath}
\usepackage{amssymb}
\usepackage{amscd}
\usepackage{wasysym}
\usepackage[ansinew]{inputenc}
\usepackage[T1]{fontenc}
\usepackage{ae,aecompl}

\newcommand{\ri}{{ \rm i }}

\newcommand{\rd}{{ \rm d }}

\newcommand{\be}{\begin{eqnarray}}
\newcommand{\ee}{\end{eqnarray}}
\newcommand{\nn}{\nonumber}

\newcommand{\ket}[1]{|#1\rangle}
\newcommand{\bra}[1]{\left\langle#1\right|}

\renewcommand{\vec}[1]{\mathbf{#1}}
\newcommand{\QQ}{\textsl {Q}}
\newcommand{\PP}{\textsl {P}}

\renewcommand{\Re}{{\rm Re}}
\renewcommand{\Im}{{\rm Im}}
\newcommand{\Dop}{\mathcal{D}}
\newcommand{\abl}[1]{\frac{\partial}{\partial #1}}

\begin{document}
\bibliographystyle{apsrev}
\title{Exact number conserving phase-space dynamics of the $M$-site Bose--Hubbard model}
\author{F. Trimborn, D. Witthaut and H. J. Korsch}
\email{korsch@physik.uni-kl.de}
\affiliation{FB Physik, Technische Universit{\"a}t Kaiserslautern,
D-67653 Kaiserslautern, Germany}
\date{\today }

\begin{abstract}
The dynamics of $M$-site, $N$-particle Bose-Hubbard systems is described in
quantum phase space constructed in terms of generalized $SU(M)$ coherent states.
These states have a special significance for these systems as they describe
fully condensed states. Based on the differential algebra developed by
Gilmore, we derive an explicit evolution equation for the (generalized)
Husimi-(\QQ)- and Glauber-Sudarshan-(\PP)-distributions. Most remarkably, these
evolution equations turn out to be second order differential
equations where the second order terms scale as $1/N$ with the particle
number. For large $N$ the evolution reduces to a (classical) Liouvillian
dynamics. The phase space approach thus provides a distinguished instrument to
explore the mean-field many-particle crossover. In addition, the thermodynamic
Bloch equation is analyzed using similar techniques.
\end{abstract}

\maketitle


\section{Introduction}
The phase space formulation of quantum mechanics is nearly as old as the theory itself \cite{Wign32}. Although the representation is equivalent to the Schr\"{o}dinger or Heisenberg picture the resemblance between the classical and the quantum phase space description reveals interesting analogies and differences between the two regimes. \\
However, the usefulness of this approach is by no means restricted to illustrations. In quantum optics there is a wide range of applications of phase space methods (for a general overview see, e.g.,~\cite{Schl01}). One particular technique which we will exploit in this paper is the association of non-commuting operator equations with c-number differential equations. In the case of the position or momentum representation this differential form of the operators is common knowledge. By the same token the correspondence between an operator acting on a density operator and a differential operator acting on a phase space distribution in flat phase space is widely used, e.g. in the context of quantum noise \cite{Gard04}. Strangely enough, these methods were for a long time restricted to the description of systems which can be described by the dynamic group of the harmonic oscillator like a spinless non-relativistic quantum particle or a mode of the quantized radiation field. Only eight years ago a general algorithm to construct an $s$-parametrized family of phase space distributions for systems with arbitrary dynamical Lie groups has been proposed \cite{Brif98b}. Therefore it has taken thirty years to extend the work of Cahill and Glauber \cite{Cahi69} and Agarwal and Wolf \cite{Agar70} from the Heisenberg-Weyl group to phase space topologies differing from the complex plane. \\
In this paper, we will present a phase space analysis of the Bose--Hubbard Hamiltonian,
\be\label{eqn-mehrmodenbosehubbard}
\hat H &=& \sum_{i=1}^M \epsilon_i \hat n_i -\Delta \sum_{i=1}^{M-1} \left(\hat a_i^\dagger \hat a_{i+1} + \hat a_{i+1} ^\dagger \hat a_i\right)\nn\\
&& \qquad + \frac U 2 \sum_{i=1}^M\left(\hat n_i(\hat n_i -1)\right),
\ee
with $\hat a_j,\hat a_j^\dagger$ being the bosonic annihilation and creation operators. This model is a paradigm for the study of strongly correlated bosonic systems, describing two apparently very different systems, Josephson junction arrays and bosons in optical lattices (see, e.g.,~\cite{Brud05} and references therein). In both cases, the parameter $U$ describes the on-site interaction between the bosons, the hopping element $\Delta$ gives the tunneling strength confined to nearest neighbors, and $\epsilon_j$ represents the chemical potential at each site $j$. In dependence of the parameter ratio, the system undergoes a quantum phase transition from a superfluid phase for $\Delta \gg U$, characterized by long range coherence and vanishing gap in the excitation spectrum, to the Mott phase for $U \gg \Delta$, dominated by localization effects \cite{Fish89b}. Especially the prediction \cite{Jaks98} and the spectacular experimental realization \cite{Grei02} of the latter system attracted a lot of interest, since this shows that optical lattices can be seen as a kind of laboratory for strongly correlated many-body systems. \\
The dynamical group of the Bose--Hubbard model for $M$ sites is spanned by the normally ordered operators $\hat a_j^\dagger \hat a_k$ with $j,k\in \{1,2,...,M\}$ and is hence equivalent to the special unitary group $SU(M)$. This is underlined by the fact that every group element as well as the Hamiltonian itself commutes with the particle number operator $\hat N=\sum \nolimits_{j=1}^M \hat a_j^\dagger \hat a_j$. Consequently an analysis in terms of the flat phase space and the use of related methods, like Glauber coherent states, is not adequate. For instance, the single operators $\hat a_j, \hat a_j^\dagger$ lead to Hilbert spaces with different particle numbers and the order parameter $\langle \hat a_j \rangle$ obviously vanishes. These facts have been taken into account by some recent approaches \cite{Buon05,Tikh06}.\\
In the present paper we will show that taking into consideration the particle number conservation explicitly has significant advantages regarding the physical interpretation and the justification of common approximations:
Since the dynamical group is no longer a direct sum of the Heisenberg--Weyl group, but given by $SU(M)$ symmetries, one has to apply an extended concept of coherent states \cite{Pere86}. These states obey a generalized minimum uncertainty relation and stay coherent under an evolution which is linear in the generators of the dynamical group. Moreover, as we will argue in this paper, the corresponding generalized coherent states are equivalent to the fully condensed states and are therefore of high physical significance. Thus, an analysis in terms of phase space distributions based on these states emphasizes directly every deviation from a product state matching a macroscopic wave function. These states are the basis of the approximate description by the discrete Gross-Pitaevskii equation, which qualifies the phase space distributions as an excellent tool to analyze and illustrate the mean--field many--particle correspondence. Furthermore, the presented method conserving the $SU(M)$ symmetry is particularly suitable to derive and justify mean-field equations and truncated phase space approaches.

The paper is organised as follows: In the next section we will recapitulate the concept of generalized coherent states and discuss the relevant cases. Here we will also show that every condensed states can be written as a $SU(M)$ coherent state and vice versa. In the third section, we will introduce a method to map operator equations onto c-number differential equations for the $SU(M)$ algebra. This technique will enable us to calculate the exact phase space dynamics for the Husimi-(\QQ)- and the Glauber-Sudarshan-(\PP)-function of the Bose-Hubbard model which is without any approximations or restrictions to the initial state given by a second order linear differential equation in the parameter space of the $SU(M)$ coherent states. A comparison to the classical Liouville equation in phase space reveals a deeper connection: The exact phase space dynamics consists of a first order differential equation plus a many--particle quantum correction of second order decaying with the particle number as $1/N$. This yields an obvious justification for a truncation of the evolution equations for large particle numbers, in contrast to established methods as the truncated Wigner approach \cite{Stee98}, where the justification is rather difficult. The first order differential terms can be thought of as a classical term since they are identical to the results of the Liouville equation. However, this technique is not restricted to dynamics. As another possible application, we will map the thermodynamical Bloch equation onto a differential equation. Finite temperature effects in the Bose--Hubbard model as, e.g., thermal fluctuations have recently attracted a lot of experimental and theoretical interest \cite{Gati06,LuYu06,Plim04}. A closer analysis shows that the resulting density matrix can be also decomposed into a classical contribution, affected only by the Gross-Pitaevskii Hamiltonian function, plus a many--particle correction. These examples show that the phase space approach is a distinguished instrument to explore the mean--field many--particle crossover. 

\section{Generalized coherent states}

The basic ingredient which we will need in the following is the concept of generalized coherent states for systems with an arbitrary dynamical Lie group \cite{Pere86}. The parameter space of the generalized coherent states determines the corresponding phase space and reflects the physical properties of the system by its geometric structure. Moreover it has been shown that one can construct explicitly a family of phase-space distributions for a system with arbitrary Lie group symmetry relaying on this concept \cite{Brif98b}. In this section we will provide the basics and the notations for the following.\\

So, let $G$ be the dynamical Lie group of the relevant quantum system. For simplicity we assume that $G$ is connected, simply connected and has a finite dimension, which is the case for the matrix Lie groups considered in this paper. It is important to note that the general approach does not rely on these assumptions. The unitary irreducible representation of the dynamical group $G$ acting on the Hilbert space will be denoted by $T$. With these preliminaries, we can define the generalized coherent states by the action of an element of the unitary irreducible representation $T$ on a fixed normalized reference state $\ket{\psi_0}$:
\be
\ket{\psi_g}=T(g)\ket{\psi_0}, \qquad g\in G.
\ee
Even though the choice of the reference state is in principle arbitrary, it influences strongly the shape of the coherent states and the structure of the corresponding phase space. Therefore a physically motivated choice would be an extremal state of the Hilbert space like the vacuum ground state for the Heisenberg-Weyl group or the lowest/highest spin state for the case of $SU(M)$. Mathematically these states correspond to the highest/lowest weight states of the unitary irreducible representation \cite{Zhang90}. \\

The isotropy subgroup or maximum stability group $H \subset G$ consists of every element which leaves the reference state invariant up to a phase factor. Formally one can write
\be
T(h)\ket{\psi_0}=e^{\ri \phi(h)} \ket{\psi_0} \quad \mbox{with} \;\phi(h)\in \mathbb{R} \;\; \forall h\in H.
\ee
With respect to the coherent states, there is a unique decomposition for every element $g \in G$ into a product of two elements, one of the isotropy subgroup $H$ and one of the coset space $G/H$:
\be
g=\Omega h, \qquad g\in G,\;h\in H\; \mbox{and}\; \Omega \in G/H.
\ee
Hence, there is a one-to-one correspondence between the elements $\Omega(g)$ of the coset space $H/G$ and the coherent states $\ket{\Omega}\equiv \ket{\psi_\Omega}$ which preserves the algebraic and topological properties. This construction guarantees the characteristic property of the coherent states: a coherent state stays coherent under a time evolution linear in the generators of the dynamical group.

Another important property we will need in the following is the (over)completeness of the coherent states \cite{Pere86}, which leads to the resolution of the identity operator of the Hilbert space,
\be
\int_{G/H} \ket{\Omega}\bra{\Omega} \rd \mu(\Omega) = I,
\ee
where $\rd \mu(\Omega)$ denotes the invariant measure on the coset space. Moreover, this fact guarantees that one can uniquely reconstruct the density matrix from the the \PP-- or \QQ--distribution.

\subsection{Glauber states}

As the Wigner function \cite{Wign32} and the Moyal quantization \cite{Moya49}, the coherent states where first introduced for the Heisenberg-Weyl algebra $h_4=\{\hat a,\hat a^\dagger,\hat a^\dagger \hat a\equiv \hat n,I\}$, with $\hat a$ and $\hat a^\dagger$ being the bosonic annihilation and creation operators. One of the first applications was the description of a mode of the quantized radiation field modeled by harmonic oscillators \cite{Glau63a}.
In this case the unitary irreducible representation of an arbitrary group element $g \in H_4$ can be decomposed as
\be
T(g)=e^{\alpha \hat a^\dagger - \alpha^* \hat a}e^{\ri(\delta \hat n + \phi I)} \quad \alpha\in \mathbb{C},\; \delta,\phi\in\mathbb{R},
\ee 
with the stability subgroup $U(1)\times U(1)$ being generated by $\{\hat n ,I\}$. Therefore the phase space is isomorphic to the complex plane $H_4/U(1)\times U(1)\cong \mathbb{C}$, parametrized by the complex parameter $\alpha$ and the typical representative of the coset space
\be
\hat D(\alpha) \equiv e^{\alpha \hat a^\dagger - \alpha^* \hat a}
\ee
is just the well--known displacement operator. With the physically motivated choice of the vacuum ground state $\ket{0}$ as the reference state one obtains the famous Glauber states 
\be
\ket{\alpha} \equiv \hat D(\alpha) \ket{0}.
\ee 

The generalization to more then one mode is straightforward, since the multimode group $\bigoplus_{i\in \mathbb{N}}\{\hat a_i,\hat a_i^\dagger,\hat a_i^\dagger \hat a_i\equiv \hat n_i,I\}$ is just a direct sum of the single--mode group. Thus the multimode Glauber states can be obtained as a direct product of the single--mode Glauber states,
\be
\ket{\boldsymbol{\alpha}} & = & \prod_{i=1}^M \ket{\alpha_i} \nn\\
& = & \prod_{i=1}^M e^{\alpha_i\hat a_i^\dagger-\alpha^*_i\hat a_i} \ket{\vec 0},
\ee
with $\ket{\vec 0}$ being the multimode vacuum ground state. Due to this factorization the well--known properties of the single--mode Glauber states can be transferred easily.

\subsection{$SU(M)$--coherent states}

In the case of the Bose-Hubbard model (\ref{eqn-mehrmodenbosehubbard}) with $M$ sites, the dynamical group is equivalent to the special unitary group $SU(M)$, spanned by the generalized angular momentum operators $\hat E_{jk}=\hat a_{j}^\dagger \hat a_k$ with $j,k\in \{1,2,...,M\}$. These fulfill the algebraic commutation relations
\be 
\left[\hat E_{jk},\hat E_{mn}\right]=\hat E_{jn}\delta_{km}-\hat E_{mk}\delta_{nj}
\ee
and conserve the particle number $\hat N = \sum \nolimits_{j=1}^M \hat E_{jj}$, since
\be
\left[\hat E_{jk},\hat N \right] = 0.
\ee
As already argued above, a suitable choice of the reference state is the maximum spin state, corresponding to the state with the entire population in the first well $\ket{N,0,\ldots,0}$. With respect to this state, an arbitrary element of the unitary irreducible representation can always be decomposed as
\be
&&T(g)\ket{N,0,\ldots,0} = \exp{\left(\sum _{k=2}^M (y_{k1}\hat E_{k1}+y_{1k}\hat E_{1k})\right)}\nn\\
&& \phantom{T}\times \exp{\left(\sum_{k,l=2}^M y_{kl}\hat E_{kl} + y_{11}\hat E_{11}\right)}\ket{N,0,\ldots,0}
\ee
into an element of the coset space and an element of the stability group $U(M-1)\times U(1)$ \cite{Gilm75}. Given that $\hat E_{jk}=\hat E_{kj}^\dagger$, we have to assume that $y_{jk}^*=y_{kj}$ in order for the argument of the exponentials to be anti-hermitian. Therefore we get the $SU(M)$ coherent states by the action of the representative of the coset space onto the reference state
\be
&&\hat{\mathcal{R}}(\vec y)\ket{N,0,\ldots,0} \nn\\
&&\qquad =\exp\bigg(\sum _{k=2}^M (y_{k1}\hat E_{k1}-y_{k1}^*\hat E_{k1}^\dagger)\bigg)\ket{N,0,\ldots,0} \nn\\
&&\qquad =: \ket{\vec y}.
\ee
The parameter space of the coherent states is spanned by the $M-1$ complex parameters $y_k\equiv y_{k1}$ with $k \in \{2,...,M\}$ of the coset space and can thus be identified with the $2(M-1)$ sphere which is topologically equivalent to $U(M)/U(M-1)\times U(1) \cong SU(M)/U(M-1)$.
Due to this analogy one can interpret the coset representative as a rotation of the reference state on the multidimensional sphere. To assure that the parametrization is unique one has to demand that the parameters are bounded as $\sum_{k= 2 }^M y_k^* y_k \leq (\pi / 2)^2$. In the case of two sites the definition of the coherent states reduces to the spin coherent states or Bloch states \cite{Arec72,Radc71}.

Anyhow, a parametrization by the $(M-1)$ independent complex parameters $(x_2,\ldots,x_M)$ of the site together with the real dependent parameter of the first site $x_1^*=x_1$ is physically more reasonable. These parameters represent the probability amplitudes at the respective sites, reflect directly the particle conservation
\be
x_1^2 + \sum_{k\geq 2}^M x_2^*x_2 =1,
\ee
and the irrelevance of the global phase. By means of the generalized Baker-Campell-Hausdorff formula one can show the relation
\be \label{eqn_ErgebnisBCH}
\hat{\mathcal{R}} \hat a_1^\dagger \hat{\mathcal{R}}^{-1}&=& \cos(\left\|y\right\|) \hat a_1^\dagger + \frac {\sin(\left\|y\right\|)}{\left\|y\right\|} \sum_{k=2}^M y_k \hat a_k^\dagger
\ee
with the abbreviation $\left\|y\right\|^2 \equiv \sum_{k=2}^M |y_k|^2$.
This leads directly to the parameter transformation
\be
x_1 = \cos(\left\|y\right\|), \qquad x_k  =  \frac{\sin(\left\|y\right\|)}{\left\|y\right\|} y_k, \qquad k \geq 2
\ee
and the representation of the $SU(M)$ coherent states in terms of the complex amplitudes $(x_1,x_2,\ldots,x_M)$:
\be
\ket{\vec y} &=& \hat{\mathcal{R}} \ket{N,0,\ldots,0}\nn\\
&=& \frac 1 {\sqrt{N!}} \hat{\mathcal{R}} \hat a_1^{\dagger N} \ket{0,0,\ldots,0}\nn\\
&=& \frac 1 {\sqrt{N!}} \left(\sum_{k=1}^M x_k \hat a_k^\dagger \right)^N \hat{\mathcal{R}} \ket{0,0,\ldots,0}\nn\\
&=& \frac 1 {\sqrt{N!}} \left(\sum_{k=1}^M x_k \hat a_k^\dagger \right)^N \ket{0,0,\ldots,0} \nn\\
&=:& \ket{\vec x}_N,
\ee
where we have used the commutation relation (\ref{eqn_ErgebnisBCH}). The last relation reveals another interesting property of the $SU(M)$ coherent states. In the case of the Bose--Hubbard model these states are equivalent to the fully condensed states, since they can always be written as a product state. This characteristic trait is certainly not trivial and it cannot be generalized to other dynamical groups since it is an intrinsic property of the $su(M)$ algebra. Moreover, this fact also singles out the physical significance of an analysis in terms of phase space distribution which are based on the $SU(M)$ coherent states.

\section{Differential algebra}

In this section we will present a formalism to map quantum observables onto differential equations acting on the continuous parameter space of the coherent states based on the ideas of Gilmore \cite{Gilm75} which we will use to calculate the exact phase space dynamics for the Bose--Hubbard model. In contrast to other approaches, for example based on the star product (see \cite{Klim02b} and references therein), this formalism is not restricted to the case of just two sites or to the special case of some dynamical groups \cite{Zhang90}. \\

\subsection{Flatland}

In the field of quantum optics the modus operandi for the Heisenberg-Weyl group $H_4$ and the Glauber coherent states is well--known (see, e.g., \cite{Gard04} and references therein). Since the Glauber states expressed in Fock states $\ket{n}$,
\be
\ket{\alpha}&=& e^{-\frac 1 2 \alpha \alpha^*}\sum_{n=0}^{\infty}\frac{\alpha^n}{\sqrt{n!}}\ket{n}\nn\\
& = & \sum_n f_n (\alpha) \ket{n},
\ee
form an overcomplete basis, one can replace the action of the bosonic creation and annihilation operators by first order linear differential equations acting on the function $f_n(\alpha) \equiv \exp(-\frac 1 2 \alpha\alpha^*) \alpha^n / \sqrt{n!}$. This yields the differential operators $\Dop^k$ acting on a ket state
\begin{align}
& \hat A \ket{\alpha}=\Dop^k (\hat A) \ket{\alpha}\nn\\
& \mbox{with}\quad \Dop^k(\hat a^\dagger)=\frac{\partial}{\partial \alpha} + \frac{1}{2} \alpha^* \quad \mbox{and} \quad \Dop^k (\hat a) = \alpha.
\end{align}
Since we are in the following interested in phase space densities corresponding to density operators and therefore to products of functions $f_n(\alpha) f_m(\alpha^*)$, we need the differential operators $\Dop^l$ acting from the left side on the coherent state projectors:
\begin{align}
& \hat A \ket{\alpha}\bra{\alpha}=\Dop^l (\hat A)\ket{\alpha}\bra{\alpha}\nn\\
& \mbox{with}\quad \Dop^l(\hat a^\dagger)=\frac{\partial}{\partial \alpha} + \alpha^* \quad \mbox{and} \quad \Dop^l(\hat a)=\alpha.
\end{align}
The generalization to operators $\Dop^r$ acting from the right, 
\be
\Dop^r(\hat A)=\left[\Dop^l(\hat A^\dagger )\right]^*,
\ee
and to multimode Glauber states is straightforward:
\begin{align}
& \Dop^l(\hat a_i^\dagger)=\frac{\partial}{\partial \alpha_i} + \alpha_i^* = \Dop^r(\hat a_i)^* \nn \\
& \Dop^l(\hat a_i)= \alpha_i = \Dop^r(\hat a_i^\dagger)^*.
\end{align}

By means of the properties of the differential operators acting on arbitrary elements of the multimode algebra $\hat A,\hat B$ with $r,s \in \mathbb{C}$,
\be \label{eqn_eigenschaftdalgebr}
& \Dop^l(r\hat A + s \hat B)=r\Dop^l(\hat A) + s\Dop^l (\hat B)\\
& \Dop^l(\hat A \hat B)=\Dop^l(\hat B)\Dop^l(\hat A)\\
& \Dop^l\left(\left[\hat A,\hat B\right]\right)=\left[\Dop^l(\hat B),\Dop^l(\hat A)\right],
\ee
one can show that the differential operators conserve the algebraic structure. Therefore the differential operators of the generators of the Heisenberg-Weyl algebra form itself a closed (differential) algebra.

\subsection{From the plane to the sphere}

The $su(M)$ algebra is generated by the set of operators $\{\hat E_{jk}=\hat a_j^\dagger \hat a_k\}$ with $j,k\in \{1,2,3,...,M\}$. In the case of the multimode Glauber states, the corresponding differential operators read
\be
\Dop^l(\hat E_{jk})=\Dop^l(\hat a_k)\Dop^l(\hat a_j^\dagger)=\alpha_k \partial _{\alpha_j} +  \alpha_k \alpha_j^*.
\ee
Using the transformation 
\be
\alpha_i=x_i \alpha e^{i\phi},\quad \alpha=\sum_i \left(\alpha_i \alpha_i^* \right)^{\frac 1 2}, \quad e^{i\phi}=\frac{\alpha_1}{|\alpha_1|},
\ee
to the $M-1$ complex parameters $\vec x=(x_2,x_3,\ldots,x_M)^t$, the norm $\alpha$ and the global phase $\phi$,
one obtains the differential form of the generalized angular momentum operator in terms of the multimode Glauber states
\be\label{eqn_Ejkdef}
&\Dop^l(\hat E_{jk})=&x_k \frac {\partial}{\partial x_j} + x_k x_j^* \left(\frac \alpha 2 \frac{\partial}{\partial \alpha}  + \alpha^2 \right) \nn \\
&&-\frac 1 2 x_k x_j^* \left( \vec x \vec \nabla + \vec x^* \vec \nabla^*\right).
\ee
Here we have used the definition
\be \label{eqn_abbrev1}
\vec x \vec \nabla + \vec x^* \vec \nabla^*=\sum_{k=2}^M x_k \abl{x_k} + x_k^* \abl{x_k^*}.
\ee
The parameter $x_1=x_1^*$ is fixed by the normalization
\be
x_1=\sqrt{1-\sum_{k=2}^M x_k^* x_k},
\ee
which leads to the following definition of the derivative with respect to the dependent parameter:
\be \label{eqn_indablmitphi}
\abl{x_1} \equiv \frac{1}{2 x_1}\left(\frac \partial{\partial (\ri \phi)} - \vec x \vec \nabla + \vec x^* \vec \nabla^* \right)\equiv -\abl{x_1^*}.
\ee

To reduce the $M$ independent complex parameters of the multimode Heisenberg--Weyl group to the $(M-1)$ independent complex variables parametrizing the $SU(M)$ coherent states, one has to invert the relation between the projectors for the multimode Glauber states and the $SU(M)$ coherent states $\ket{\vec x}_{N}$:
\be
\ket{\alpha}\bra{\alpha} = \sum_{L,N=0}^\infty e^{- |\alpha|^2} \frac{\alpha^{N+L} e^{\ri\phi(N-L)}}{\sqrt{N!L!}} \ket{\vec x} _{N}\bra{\vec x} _{L}\,.
\ee
This can be done using the following homomorphism \cite{Gilm75}
\be
\lim_{\alpha^2\rightarrow 0} \left(\frac{\partial}{\partial \alpha}\right)^N e^{- \alpha^2} \oint \ket{\alpha}\bra{\alpha} \frac{\rd \phi}{2\pi} = \ket{\vec x}_{N}\bra{\vec x}_{N}
\ee
and the relation
\be
\left(\frac \alpha 2 \frac{\partial}{\partial \alpha}  + \alpha^2 \right) e^{-\alpha^2}\frac{\alpha^{2N}}{N!}=N e^{-\alpha^2}\frac{\alpha^{2N}}{N!}.
\ee 
A short calculation gives the desired result
\be
&& \lim_{\alpha^2\rightarrow 0} \left(\frac{\partial}{\partial \alpha}\right)^N e^{- \alpha^2} \oint \Dop^l(\hat a_j^\dagger \hat a_k) \ket{\alpha}\bra{\alpha} \frac{\rd \phi}{2\pi} \nn\\
& = & x_k \frac {\partial}{\partial x_j} + x_k x_j^* \left( N -\frac 1 2 \left( \vec x \vec \nabla + \vec x^* \vec \nabla^*\right) \right) \ket{\vec x}_{N}\bra{\vec x}_{N}\nn\\
& \equiv & \Dop^l(\hat E_{jk})\ket{\vec x}_{N}\bra{\vec x}_{N},
\ee
where we used the following abbreviation:
\be \label{eqn_abbrev2}
\abl{x_1} \equiv - \frac{1}{2 x_1}\left(\vec x \vec \nabla - \vec x^* \vec \nabla^* \right)\equiv -\abl{x_1^*}.
\ee
A comparison to equation (\ref{eqn_indablmitphi}) shows that the differentiation no longer depends on the global phase. This can be understood as an averaging effect of the integration over the angle $\phi$, which is part of the homomorphism.

\section{Dynamics}

\subsection{The Husimi--distribution}

The time evolution of the Husimi-- or \QQ--distribution,
\be
\QQ(\Omega)=\bra{\Omega}\hat \rho \ket{\Omega},
\ee
(with $\ket{\Omega}$ being the generalized coherent states for the relevant symmetry group) follows from the formal time dependence of the density operator
\be
\dot{\hat{\rho}} = - \frac \ri \hbar \left[\hat H, \hat \rho \right] = - \frac \ri \hbar \hat H \hat \rho + \frac \ri \hbar \hat \rho \hat H.
\ee
By means of the relation
\be
\frac {\partial}{\partial t}  \QQ (\Omega,t)= {\rm tr} (\dot{\hat \rho} \ket{\Omega}\bra{\Omega}),
\ee
the properties of the trace and the hermiticity of the Hamiltonian one finds
\be
\frac {\partial}{\partial t}  \QQ (\Omega,t) & = & \frac  \ri \hbar \left( \Dop^l (\hat H)-\Dop^l (\hat H)^* \right) \QQ (\Omega,t)\nn \\
&=& - \frac  2 \hbar \,\Im \, \left(\Dop^l (\hat H)\right) \QQ (\Omega,t),
\ee
independent of the specific structure of the dynamical group. In the following we will use rescaled units with $\hbar =1$.\\

To evaluate the imaginary part of the differential operator $\Dop^l (\hat H)$  for the Bose--Hubbard Hamiltonian (\ref{eqn-mehrmodenbosehubbard}),
\be
\frac {\partial \QQ (\vec x,t)}{\partial t} &=& -  2 \; \Im  \biggl( \sum_{i=1}^M \epsilon_i \Dop^l (\hat n_i)  +\frac U 2 \sum_{i=1}^M \left(\Dop^l(\hat n_i)^2\right) \\
&&-\Delta \sum_{i=1}^{M-1} \left(\Dop^l(\hat a_i^\dagger \hat a_{i+1})+ \Dop^l(\hat a_{i+1} ^\dagger \hat a_i)\right) \biggl) \QQ (\vec x,t), \nn
\ee
we change once again the parametrization by an amplitude phase decomposition:
\be \label{eqn_parameterpq}
x_1=\sqrt{p_1},\qquad x_i=\sqrt{p_i}e^{-\ri q_i} \qquad 2 \leq i \leq M.
\ee
In the case of the Bose--Hubbard model, the $p_j$ refer to the relative occupation in the $j$-th well and $q_j$ describes the relative phase between the $j$-th and the first well.

Since the results for the differential operators can be used for every Hamiltonian whose dynamical symmetries are a $SU(M)$ group, the explicit form of the differential operators may be of general interest. A lengthy calculation yields the results for the differential operators for $j=1$,
\be
&& \Im \left(\Dop^l (\hat n_1)\right) = - \frac 1 2 \sum_{k\geq 2} \abl{q_k}\\
&& \Im \left(\Dop^l (\hat n_1)^2\right) = p_1 \left(\sum_{k,k'\geq 2} p_k \frac{\partial^2}{\partial p_k \partial q_{k'}}-N\sum_{k\geq 2} \abl{q_k}\right)\nn\\
&& \Im \left( \Dop^l(\hat a_1^\dagger \hat a_{2})+ \Dop^l(\hat a_{2} ^\dagger \hat a_1)\right) = - \frac 1 2 \cos q_2 \sqrt{\frac {p_2} {p_1}}\sum_{k\geq2}\abl{q_k} \nn\\
&& \qquad +\sqrt{p_1 p_2}\sin q_2 \abl{p_2} +\frac 1 2 \cos q_2 \sqrt{\frac {p_1} {p_2}}\abl{q_2}, \nn
\ee
and for $2 \leq j\leq M$,
\be
&& \Im \left(\Dop^l (\hat n_j)\right)= \frac 1 2 \abl{q_j}\\
&& \Im \left(\Dop^l (\hat n_j)^2\right) = p_j\left(N-\sum_{k\geq2}^M p_k\abl{p_k} \right)\abl{q_j}+p_j\frac{\partial^2}{\partial p_j \partial q_j}\nn\\
&& \Im \left( \Dop^l(\hat a_j^\dagger \hat a_{j+1})+ \Dop^l(\hat a_{j+1} ^\dagger \hat a_j)\right) \nn\\
&&= \sqrt{p_j p_{j+1}} \sin(q_j-q_{j+1})\left(\abl{p_j}-\abl{p_{j+1}} \right) \nn\\
&&\phantom{=}  + \frac 1 2 \cos (q_j - q_{j+1}) \left(\sqrt{\frac{p_{j+1}}{p_j}} \abl{q_j} + \sqrt{\frac{p_{j}}{p_{j+1}}} \abl{q_{j+1}}\right)\nn.
\ee

The advantage of this result is that it directly gives the exact phase space dynamics of the Bose--Hubbard model in terms of the Husimi--function:
\be \label{eqn_husmisres}
&& \!\!\!\!\! \frac{\partial \QQ}{\partial t} (\vec p,\vec q,t) =  \bigg\{ \Delta \bigg(+ 2 \sqrt{p_{2} p_1} \sin q_2 \partial_{p_2} \nn\\
&& +2 \sum_{k=2}^{M-1} \sqrt{p_{k+1} p_k} \sin (q_k - q_{k+1})\left(\partial_{p_k}-\partial_{p_{k+1}}\right)\nn\\ 
&& + \sum_{k=1}^{M-1} \cos(q_{k+1}-q_k) \big(\sqrt{\frac{p_k}{p_{k+1}}} \partial_{q_{k+1}}+\sqrt{\frac{p_{k+1}}{p_k}} \partial_{q_k}\big)\bigg)\nn \\
&&+ U \bigg(N\sum_{k=2}^M(p_1-p_k)\partial_{q_k} - \sum_{k=2}^M p_k \partial_{p_k} \partial_{q_k} \nn \\
&&+ \sum_{k,k'= 2}^M (p_k-p_1)p_{k'} \partial_{p_{k'}}\partial_{q_k}\bigg) \nn\\
&&+ \sum_{k=2}^M(\epsilon_1-\epsilon_k)\abl{q_k}\;\bigg\}\;\QQ (\vec p, \vec q,t) ,
\ee
with the definitions 
\be \label{eqn_defqindependant}
q_1\equiv0, \qquad\qquad \abl{q_1}\equiv-\sum_{k=2}^M \abl{q_k}.
\ee
Therefore we have derived an explicit formula without any approximations. Before we analyze this formula we will derive the analogues result for the \PP--function.

\subsection{The Glauber--Sudarshan distribution}

The Glauber--Sudarshan or \PP--distribution is the diagonal representation of the density matrix in the basis of the generalized coherent states $\ket{\Omega}$:
\be
\hat \rho = \int \PP(\Omega) \ket {\Omega} \bra{\Omega} \rd \mu (\Omega).
\ee
Since the basis is overcomplete, this description is always possible, but not necessarily unique.\\

The differential operators for this phase space distribution, denoted below by $\tilde \Dop$ to avoid confusion, arise from a simple integration by parts of the differential operators for the Husimi--distribution:
\be
&&\int  \PP(\Omega)\Dop^l(\hat A) \ket{\Omega}\bra{\Omega} \rd \mu(\Omega)\\
&& \phantom{\int \PP(\Omega)}= \int \tilde{\Dop^l}(\hat A) \PP(\Omega) \ket{\Omega}\bra{\Omega}\rd \mu(\Omega).\nn
\ee

Thus, one can calculate the time evolution of the \PP-function using the differential operators in an analogous way as for the Husimi--distribution:
\be
\dot{\hat \rho}&=& \int \dot \PP(\Omega) \ket{\Omega}\bra{\Omega}\rd \mu(\Omega) \\
&=& \ri \int \left(\tilde \Dop^l(\hat H)^* - \tilde \Dop^l(\hat H) \right) \PP(\Omega) \ket{\Omega}\bra{\Omega}\rd \mu(\Omega),\nn
\ee
or briefly
\be
\frac{\partial}{\partial t} \PP(\Omega,t) &=& - 2 \,  \Im \,\left(\tilde \Dop^l (\hat H)\right)\, \PP(\Omega,t).
\ee

Therefore we can derive the expression for the differential operator of the generalized angular momentum operator by an integration by parts:
\be
& \tilde \Dop^l (\hat E_{jk})=&-x_k \abl{x_j} - \delta_{jk} \\
&&+ x_kx_j^*\bigg( (N+M)+\frac 1 2 (\vec x \vec \nabla + \vec x^* \vec \nabla^*)\bigg).\nn
\ee 
In this equation we used the same definitions as above in equation (\ref{eqn_abbrev2}). The origin of the minor changes compared to the case of the \QQ--function is clear: the additional factor $M$ and the $\delta$-symbol result from the different operator ordering and the sign is due to the integration by parts.

Now we can calculate the exact dynamics of the \PP--function for the Bose--Hubbard model with $M$ sites:
\be \label{eqn_PPres}
&& \!\!\!\!\! \frac{\partial \PP}{\partial t} (\vec p,\vec q,t) = \bigg\{ + \Delta \bigg(2 \sqrt{p_{2} p_1} \sin q_2 \partial{p_2} \nn\\
&& +2 \sum_{k=2}^{M-1} \sqrt{p_{k+1} p_k} \sin (q_k - q_{k+1})\left(\partial{p_k}-\partial{p_{k+1}}\right) \nn\\
&& + \sum_{k=1}^{M-1} \cos(q_{k+1}-q_k) \big(\sqrt{\frac{p_k}{p_{k+1}}} \partial{q_{k+1}}+\sqrt{\frac{p_{k+1}}{p_k}} \partial{q_k}\big)\bigg)\nn \\
&&+ U \bigg((N+M)\sum_{k=2}^M(p_1-p_k)\partial{q_k} + \sum_{k=2}^M p_k \frac{\partial^2}{\partial p_k \partial q_k}\nn\\
&&- \sum_{k,k'= 2}^M (p_k-p_1)p_{k'} \partial_{p_{k'}}\partial_{q_k}  \bigg)\bigg\}\nn \\
&&+ \sum_{k=2}^M(\epsilon_1-\epsilon_k)\partial{q_k}\;\bigg\}\;\PP (\vec p, \vec q,t) ,
\ee
where we used rescaled units $\hbar=1$ and the same definitions (\ref{eqn_defqindependant}) as above.

A comparison with the result for the Husimi--distribution (\ref{eqn_husmisres}) shows that due to the operator ordering the interaction strength now varies with the particle number plus the number of sites, $U(N+M)$. Apart from this issue, the first order differential form is exactly the same. The second order contribution has apparently the same structure as above, but the sign has changed. In both cases the second order term vanishes in the macroscopic limit $N\rightarrow \infty$ with $UN$ fixed as $\mathcal{O}(1/N)$.

\subsection{Liouville dynamics}\label{abschn_Liou}

In the mean-field limit, the dynamics of a BEC in an optical lattice is given by the celebrated discrete Gross-Pitaevskii equation (GPE) or discrete nonlinear Schr\"{o}dinger equation (see, e.g.,~\cite{Trom01} and references therein):
\be
\ri \dot x_j = \epsilon_j x_j-\Delta (x_{j+1} + x_{j-1}) + UN |x_j|^2 x_j.
\ee
Using again the decomposition into amplitude and phase (\ref{eqn_parameterpq}), the dynamics can be reformulated as classical Hamiltonian equations
\be
\dot q_i = \frac{\partial \mathcal{H}}{\partial p_i}, \qquad \dot p_i = - \frac{\partial \mathcal{H}}{\partial q_i},
\ee
with the corresponding Hamiltonian function
\be \label{eqn_gpe}
\mathcal{H}(\vec p,\vec q) &=& - 2 \Delta \sum_{k=1}^{M-1} \sqrt{p_k p_{k+1}} \cos(q_{k+1}-q_k) \nn\\
&& + \frac {UN}2 \sum_{k=1}^M p_k^2 + \sum_{k=1}^M \epsilon_k p_k.
\ee
One should keep in mind that the parameters of the first well are not independent. The GPE describes the exact dynamics for vanishing interaction $U\equiv 0$ and an initially coherent state, since then an initial state stays coherent and the description by a single particle density matrix contains no approximations. 

A classical phase space distribution $\rho(\vec p,\vec q,t)\rd \vec p \rd \vec q$, with $\vec p, \vec q$ being canonical conjugate variables, describes the probability that an ensemble of particles will be found in an infinitesimal phase space element $\rd \vec p \rd \vec q$. The dynamics under the Hamiltonian function $\mathcal{H}$ is governed by the classical Liouville equation
\be
\frac{\rd \rho}{\rd t}=\frac{\partial \rho}{\partial t} + \{\rho, \mathcal{H}\} = 0,
\ee
where $\{\cdot,\cdot\}$ denotes the classical Poisson bracket.
The resulting evolution equations for the Hamiltonian function (\ref{eqn_gpe}) are
\be
\frac{\partial \rho}{\partial t}&=&  \sum_{k=2}^M \frac{\partial \mathcal{H}}{\partial q_k} \frac{\partial \rho}{\partial p_k} - \sum_{k=2}^M \frac{\partial \mathcal{H}}{\partial p_k} \frac{\partial \rho}{\partial q_k} \\
&=& +2\Delta \sqrt{p_{2} p_1} \sin q_2 \partial_{p_2} \rho \nn\\
&& +2 \Delta \sum_{k=2}^{M-1} \sqrt{p_{k+1} p_k} \sin (q_k - q_{k+1})\left(\partial_{p_k}\rho-\partial_{p_{k+1}}\rho \right)\nn\\ 
&& + \Delta \sum_{k=1}^{M-1} \frac{\cos(q_{k+1}-q_k)}{\sqrt{p_k p_{k+1}}} \big(p_k \partial_{q_{k+1}} \rho + p_{k+1} \partial_{q_k} \rho\big)\nn \\
&&+ UN\sum_{k=2}^M(p_1-p_k)\partial_{q_k}\rho + \sum_{k=2}^M(\epsilon_1-\epsilon_k)\abl{q_k}\rho. \nn
\ee
A comparison with (\ref{eqn_husmisres}) and (\ref{eqn_PPres}) shows that the exact phase space dynamics consists of a first order differential equation plus a many--particle quantum correction of second order vanishing in the macroscopic limit $N\rightarrow \infty$ with $UN$ fixed. The first order terms can be thought of as the classical evolution since they are identical to the results of the Liouville equation. Thus, in the noninteracting case this result coincides with the many--particle result -- the Liouville equation is exact. In this case the GPE describes the evolution of the center or maximum of the phase space distribution.\\
However, the description in quantum phase space goes beyond the area of validity of the GPE since there are no restrictions on the shape of the initial state, up to the usual ones set up by the uncertainty relation. In the interacting case the first order part is reproduced by the classical Liouville equation, but without the term depending on the operator ordering. The first order interaction term is responsible for a variation of the shape of the state, therefore an initial state stays no longer coherent. This fact is usually denoted as the \textsl{break--down of mean--field} \cite{Vard01b,Vard01a,Cast97}, indicating that the description by a single mean--field trajectory corresponding to the evolution of the center of the coherent state is no longer valid. Indeed this breakdown is resolved by using the Liouville approach, where we can take into account the variation of the shape of the initial state and therefore effects due to variation of the higher moments. The second order differential corrections to the classical Liouville equation decay with increasing particle number as $1/N$ in the macroscopic limit. These terms are responsible for many--particle effects as tunneling in quantum phase space and (self-)interference. It is interesting to note that both the Liouville equation and the whole equation without approximations conserve the normalization. In a sequel article we will illustrate the methods presented here and discuss possible applications \cite{07phaseappl}.

\subsection{Expectation values}

The expectation value of an arbitrary operator $\hat B$ in terms of \QQ-- and \PP--functions is given by the statistical average of the phase space distribution
\be
\langle \hat B \rangle &=& \int \PP_{\hat B}(\Omega) \QQ (\Omega)  \rd \mu(\Omega) \nn \\
&=& \int \PP (\Omega) \QQ_{\hat B}(\Omega) \rd \mu(\Omega), \label{eqn_erqrtPQ}
\ee
where $\PP_{\hat B}(\Omega)$ and $\QQ_{\hat B}(\Omega)$ denotes the (anti)--normally ordered Weyl-symbol of the operator $\hat B$
\be
\hat B &\equiv& \int \PP_{\hat B}(\Omega) \ket{\Omega} \bra{\Omega}\rd \mu(\Omega) \\
\QQ_{\hat B}(\Omega) &\equiv& \bra{\Omega} \hat B \ket{\Omega}.
\ee
In contrast to the symmetrically ordered Wigner function the expectation values cannot be expressed in terms of one phase space distribution alone. However, the differential algebra formalism allows also to calculate the expectation values in terms of the \QQ-function and the differential operator without using the \PP-representation and vice versa:
\be
\langle \hat B \rangle &=& {\rm Tr} (\hat B \hat \rho) \nn\\
&=&  {\rm Tr} \left( \int \hat B \ket{\Omega}\bra{\Omega} \hat \rho \rd \mu(\Omega )\right) \nn\\
&=& \int \Dop^l (\hat B) \QQ (\Omega) \rd \mu(\Omega)\nn\\
&=& \int \tilde \Dop^l (\hat B) \PP (\Omega) \rd \mu(\Omega). \label{eqn_erwartQDop}
\ee
Note the interesting correspondence between equation (\ref{eqn_erqrtPQ}) and equation (\ref{eqn_erwartQDop}) which reveals the close connection between the differential operators and the Weyl-symbols of the operator $\hat B$.

As an example, we calculate the expectation value of the generalized angular momentum operators $\hat E_{jk}=\hat a_{j}^\dagger \hat a_k$ which span the $su(M)$ algebra in the \QQ-representation:
\be
\frac 1 N \langle \hat E_{jk} \rangle &=& \int x_k x_j^* \QQ(\vec x) \rd \mu(\vec x)  \nn\\
&& + \frac 1 N \int x_k \abl{x_j} \QQ(\vec x) \rd \mu(\vec x) \nn\\
&& - \frac 1 {2N} \int x_kx_j^* (\vec x \vec \nabla + \vec x^* \vec \nabla^*) \QQ(\vec x) \rd \mu(\vec x) \nn\\
&=& \int x_k x_j^* \QQ(\vec x) \rd \mu(\vec x) + \mathcal{O}(\frac 1 N).
\ee
and by using the \PP-function:
\be
\frac 1 N \langle \hat E_{jk} \rangle &=& \frac{N+M}{N} \int x_k x_j^* \PP(\vec x) \rd \mu(\vec x)  \nn\\
&& - \frac{\delta_{jk}}{N} \int \PP(\vec x) \rd \mu(\Omega) -\frac 1 N \int x_k \abl{x_j} \PP(\vec x) \rd \mu(\vec x) \nn\\
&& + \frac 1 {2N} \int x_kx_j^* (\vec x \vec \nabla + \vec x^* \vec \nabla^*) \PP(\vec x) \rd \mu(\vec x) \nn\\
&=& \int x_k x_j^* \PP(\vec x) \rd \mu(\vec x) + \mathcal{O}(\frac 1 N).
\ee
At the first sight, the expectation values can be decomposed into the classical statistical average and a quantum many--particle correction that vanishes if the particle number $N$ becomes macroscopically large. 
Moreover, we can even concretise the result using an integration by parts and the periodic boundary conditions. This provides the following result for the \QQ-function
\be
\langle \hat E_{jk} \rangle &=& (N+M) \int x_k x_j^* \QQ(\vec x) \rd \mu(\vec x)-\delta_{jk},
\ee
and the subsequent outcome for the \PP-function:
\be
\langle \hat E_{jk} \rangle &=& N \int x_k x_j^* \PP(\vec x) \rd \mu(\vec x).
\ee
The differences are of course due to the operator ordering. For a coherent state $\ket{\vec x_0}$ with $P(\vec x)=\delta(\vec x- \vec x_0)$ we obtain 
\be
\langle \hat E_{jk} \rangle &=& N x_{k,0} x_{j,0}^*,
\ee
as expected.

The calculation of the expectation value of the generators of the $su(M)$ algebra by a classical phase space average is thus not only a good approximation for large particle numbers, but exact. Therefore the only error of the expectation values calculated using the Liouville dynamics discussed in section \ref{abschn_Liou} is caused by the truncation of the evolution equations. This error vanishes for arbitrary initial states as $\mathcal{O}(1/N)$ in the macroscopic limit $N \rightarrow \infty$ with $UN$ fixed.

\section{Thermodynamics}

The method presented above is not restricted to an analysis of the time dependence of the system, there are multifarious applications. As an example, we consider the (unnormalized) density operator of the canonical ensemble,
\be
\hat \rho=e^{-\beta \hat H}
\ee
with $\beta= 1/kT$ and $\hbar=1$.
This expression describes the quantum mechanical version of the canonical partition function in statistical mechanics obeying the Bloch equation
\be \label{eqn_bloch}
\frac{\partial \hat \rho}{\partial \beta} = -\frac 1 2(\hat \rho \hat H+\hat H \hat \rho).
\ee

\subsection{Thermodynamics of the \QQ--function}

Translating the relation (\ref{eqn_bloch}) into differential operators acting on phase space densities, namely the \QQ-function, yields the formal result:
\be
\frac{\partial \QQ }{\partial \beta} = - \Re \, \left( \Dop^l(\hat H) \right)\QQ.
\ee
Analogously to the calculations in the previous section, we can evaluate the real part in the parametrization of the relative amplitudes and phases (\ref{eqn_parameterpq}):
\be \label{eqn_qqthermo}
&&\frac{\partial \QQ (\vec p, \vec q)}{\partial \beta} = \bigg\{- N \sum_{k=1}^M \epsilon_k p_k \nn\\
&&\quad + \sum_{k=1,k'=2}^M \epsilon_k p_k p_{k'}\partial_{p_{k'}} - \sum_{k=2}^M \epsilon_k p_k \partial_{p_k} \nn\\
&& \quad + \Delta \sum_{k=1}^{M-1} \bigg(2\sqrt{p_k p_{k+1}}\cos(q_{k+1}-q_k)\nn \\
&& \quad \phantom{\Delta \sum_{k=1}^{M-1} \bigg(2} \times (N-\sum_{k'=2}^M p_{k'}\partial_{p_{k'}}+\frac 1 2 \partial_{p_{k+1}}) \bigg)\nn\\
&& \quad +\frac \Delta 2 \sum_{k=1}^{M-1} \sqrt{\frac{p_k}{p_{k+1}}} \sin(q_{k+1}-q_k)(\partial_{q_k}-\partial_{q_{k+1}}) \nn\\
&& \quad + \Delta \sum_{k=2}^{M-1} \sqrt{p_k p_{k+1}} \cos(q_{k+1}-q_k) \partial_{p_k} \nn\\
&& \quad - \frac {UN(N-1)} 2\sum_{k=1}^M p_k^2 + U(N-1)\sum_{k=1}^M p_k^2 \sum_{k'=2}^M p_{k'} \partial_{p_{k'}} \nn\\
&& \quad - U \sum_{k=2}^M p_k^2 \big((N-1)-\sum_{k'=2}^M p_{k'}\partial_{p_{k'}}\big)\partial_{p_k} \nn\\
&& \quad - \frac U 2 \sum_{k=1}^M p_k^2 \sum_{k',k''=2}^M p_{k'}p_{k''}\partial_{p_{k'}}\partial_{p_{k''}}\nn\\
&& \quad - \frac U 2 \sum_{k=2}^M p_k^2 \partial^2_{p_k} + \frac U 8 \sum_{k=2}^2 \partial^2_{q_k}\nn\\
&& \quad + \frac U 8 \sum_{k,k'=2}^M \partial_{q_k}\partial_{q_{k'}} \bigg\} \QQ (\vec p, \vec q).
\ee
Besides the lengthy expression one already recognizes an underlying structure: The leading terms of each contribution show a close analogy to the GPE Hamiltonian function (\ref{eqn_gpe}). Before we have a closer look at the connection to the classical result, we derive an expression for the solution of the Bloch equation in terms of the Glauber--Sudarshan distribution.

\subsection{Thermodynamics of the \PP--function}

Analogously to the case of the Husimi--distribution one can derive the result for the \PP-function,
\be
\frac{\partial \PP }{\partial \beta} = - \Re \, \left( \tilde{\Dop^l}(\hat H) \right)\PP.
\ee
Here the evaluation of the real part yields
\be \label{eqn_ppthermo}
&&\frac{\partial \PP (\vec p, \vec q)}{\partial \beta} = \bigg\{- (N+M) \sum_{k=1}^M \epsilon_k p_k + \sum_{k=1}^M \epsilon_k\nn\\
&&\quad - \sum_{k=1,k'=2}^M \epsilon_k p_k p_{k'}\partial_{p_{k'}} + \sum_{k=2}^M \epsilon_k p_k \partial_{p_k} \nn\\
&& \quad - \Delta \sum_{k=1}^{M-1} \bigg(2\sqrt{p_k p_{k+1}}\cos(q_{k+1}-q_k)\nn \\
&& \quad \phantom{\Delta \sum_{k=1}^{M-1} \bigg(2} \times ((N+M)-\sum_{k'=2}^M p_{k'}\partial_{p_{k'}}+\frac 1 2 \partial_{p_{k+1}}) \bigg)\nn\\
&& \quad -\frac \Delta 2 \sum_{k=1}^{M-1} \sqrt{\frac{p_k}{p_{k+1}}} \sin(q_{k+1}-q_k)(\partial_{q_k}-\partial_{q_{k+1}}) \nn\\
&& \quad - \Delta \sum_{k=2}^{M-1} \sqrt{p_k p_{k+1}} \cos(q_{k+1}-q_k) \partial_{p_k} \nn\\
&& - \frac{U(N+M)(N+M+1)}{2}\sum_{k=1}^M p_k^2 + U(2N+M) \nn\\
&& - 2U \sum_{k=2}p_k \partial_{p_k} - U (N+M+1)\sum_{k=1,k'=2}^M p_k^2 p_{k'} \partial_{p_{k'}} \nn\\
&& +2U \sum_{k=1,k'=2}^M p_k p_{k'} \partial_{p_{k'}} - \frac U 2 \sum_{k=1}^M p_k^2 \sum_{k',k''=2}^M p_{k'}p_{k''}\partial_{p_{k'}}\partial_{p_{k''}}\nn\\
&& + U \sum_{k=2}^M p_k^2 \sum_{k'=2}^M p_{k'} \partial_{p_{k'}}\partial_{p_{k}} \nn\\
&& \quad - \frac U 2 \sum_{k=2}^M p_k^2 \partial^2_{p_k} + \frac U 8 \sum_{k=2}^2 \partial^2_{q_k}\nn\\
&& \quad + \frac U 8 \sum_{k,k'=2}^M \partial_{q_k}\partial_{q_{k'}} \bigg\}\PP (\vec p, \vec q).
\ee
Note the subtle, however important differences due to operator ordering compared with equation (\ref{eqn_qqthermo}).

\subsection{Classical vs. quantum statistical mechanics}

The distribution function of the classical canonical ensemble 
\be
\rho = e^{-\beta N \mathcal{H}}
\ee
given in terms of the Hamiltonian function (\ref{eqn_gpe}) solves the Bloch equation
\be
\frac{\partial \rho}{\partial \beta} &=& -\mathcal{H} \rho \nn\\
&=&  \bigg\{ 2 \Delta N \sum_{k=1}^{M-1} \sqrt{p_k p_{k+1}} \cos(q_{k+1}-q_k)\nn\\
&& \quad -\frac {UN^2}2 \sum_{k=1}^M p_k^2 - N \sum_{k=1}^M \epsilon_k p_k \bigg\} \rho.
\ee

A comparison with equation (\ref{eqn_qqthermo}) and equation (\ref{eqn_ppthermo}) shows that the quantum many particle Bloch equation and its formulation in terms of the \QQ-- and \PP--function can also be separated into a classical contribution, the leading order of $N$, which is governed by the Gross--Pitaevskii Hamiltonian function and quantum corrections. Amongst others these additional quantum terms ensure the minimal uncertainty for low temperatures. The high temperature limit is in both cases given by an uniform distribution.

\section{Conclusion and Outlook}
In this paper we have developed phase space techniques which provide
an alternative tool to investigate and analyze the dynamics of
one-dimensional $M$-site, $N$-particle Bose-Hubbard systems.
The quantum phase space is constructed in terms of generalized $SU(M)$ 
coherent states, which conserve the number of particles. This changes the
corresponding phase space to a compact manifold. 
In the context of Bose-Einstein condensates, the $SU(M)$ coherent states 
have a special significance for these systems 
as they describe fully condensed states. 

The phase space dynamics can be treated efficiently in terms of
the differential algebra developed by Gilmore. In this way the 
$su(M)$ operator algebra is mapped onto differential operators acting
on the multimode coherent states. The resulting evolution equations 
for the (generalized) Husimi (\QQ) and Glauber-Sudarshan (\PP) 
phase space distributions are second order differential equations.
These (exact) evolution equations provide a convenient starting
point for further developments. 

Firstly, it is immediately observed that
the second order terms scale as $1/N$ and therefore vanish in the macroscopic limit $N\rightarrow \infty$ with $UN$ fixed.
For large $N$, the evolution reduces to first order equations of the
form of (classical) Liouvillian dynamics. 
The phase space approach therefore provides a remarkable direct
derivation of the celebrated many-particle mean-field limit.

Secondly, this phase space method offers a clue to generalize the 
mean-field approximation, which describes strongly localized quantum 
states by a single point in phase space. Arbitrary quantum states
can be represented by an ensemble of phase space trajectories, which is 
constructed to approximate the initial quantum phase space (Husimi)
distribution. Then each trajectory follows the (classical) mean-field 
equations. This allows a straightforward computation of expectation 
values.

Thirdly, the resulting second-order partial differential can be attacked
directly by numerical methods. 

Finally, there is the challenge to explore the regime between the
(classical) mean-field description and the full quantum dynamics
by generalizing the semiclassical phase space methods developed during
the last decades for the flat space to systems with $SU(M)$
symmetries and a compact phase space. 

In addition, it should also be noted that the evolution equations can
also be generalized to master equations describing systems coupled to an
environment or systems with an effective decay. 
In future work we will address some of these problems, starting with
first applications to the two-mode 
Bose-Hubbard system in a forthcoming article \cite{07phaseappl}.


\begin{acknowledgments}
Support from the Studienstiftung des deutschen Volkes and the Deutsche
Forschungsgemeinschaft via the Graduiertenkolleg ``Nichtlineare Optik
und Ultrakurzzeitphysik'' is gratefully acknowledged.
\end{acknowledgments}


\end{document}